\documentclass{mn2e}
\usepackage{epsfig}

\title[The Systemic Velocity of Eta Carinae]{The Systemic Velocity of
Eta Carinae}

\author[N.\ Smith]{Nathan Smith\thanks{Email:
nathans@casa.colorado.edu}\thanks{Hubble Fellow} \\ Center for
Astrophysics and Space Astronomy, University of Colorado, 389 UCB,
Boulder, CO 80309, USA}

\date{Accepted 0000, Received 0000, in original form 0000}
\pagerange{\pageref{firstpage}--\pageref{lastpage}} \pubyear{2002}

\def\arcdeg{\degr}

\begin{document}
\label{firstpage}
\maketitle
\begin{abstract}

High-resolution spectra of molecular hydrogen in the Homunculus nebula
allow for the first direct measurement of the systemic velocity of
$\eta$~Carinae.  Near-infrared long-slit data for H$_2$ 1-0~S(1)
$\lambda$21218 obtained with the Phoenix spectrometer on the Gemini
South telescope give $V_{\rm sys}$=$-$8.1$\pm$1 km s$^{-1}$
(heliocentric), or $V_{\rm LSR}$=$-$19.7$\pm$1 km s$^{-1}$, from the
average of the near and far sides of the Homunculus.  This measurement
considerably improves the precision for the value of $-$7$\pm$10 km
s$^{-1}$ inferred from neighboring O-type stars in the Carina nebula.
New near-infrared spectra also provide a high-resolution line profile
of [Fe~{\sc ii}] $\lambda$16435 emission from gas condensations known
as the Weigelt objects without contamination from the central star,
revealing a line shape with complex kinematic structure.  Previously,
uncertainty in the Weigelt knots' kinematics was dominated by the
adopted systemic velocity of $\eta$~Car.

\end{abstract}


\begin{figure}
\epsfig{file=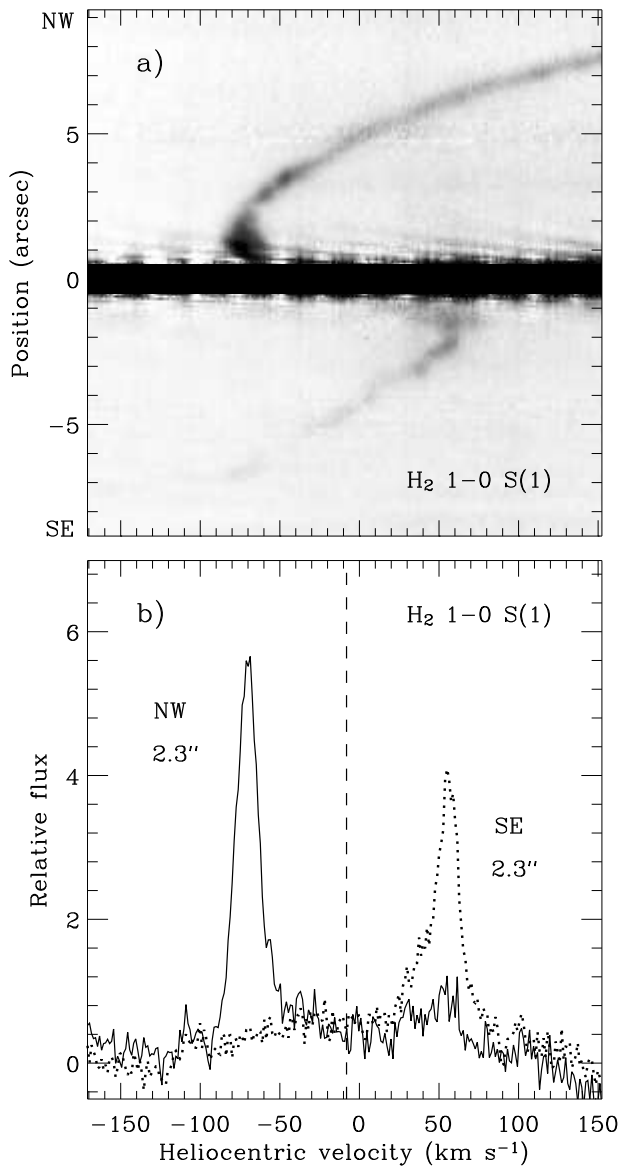,width=3.2in}
\caption{(a) Long-slit kinematics of molecular hydrogen emission in
$\eta$ Car.  The thin filament extending to the northwest emits from
the near side of the NW polar lobe of the Homunculus, while the
fainter thin filament extending away from the star to the southeast
emits from the far side of the SE polar lobe.  Continuum emission has
been suppressed for display here.  (b) Extracted spectra from
2$\farcs$3 on either side of the star; the northwest tracing is shown
with a solid line, and the southeast tracing with a dotted line.  The
measured value for $V_{\rm sys}$=$-$8.1 km s$^{-1}$ is marked with a
vertical dashed line.}
\end{figure}

\section{INTRODUCTION}

The massive star $\eta$ Carinae is currently the subject of intensive
spectroscopic study, especially with high-resolution techniques.  An
accurate and precise value for the systemic velocity is needed to
interpret line profiles of the central star and to study the
kinematics and geometry derived from emission lines in its
circumstellar ejecta.  This is especially relevant for UV echelle data
with extremely high spectral resolution (e.g., Gull et al.\ 2004), and
for any attempts to decipher orbital reflex motion if $\eta$ Car is
indeed an interacting binary system (Damineli et al.\ 2000).  Yet, the
radial velocity of the $\eta$ Car system has been notoriously
difficult to measure, because the bright central star has only broad
emission lines formed in the stellar wind, and narrow emission lines
arise only in spatially distinct circumstellar ejecta that are moving
with respect to the star system.

Davidson et al.\ (1997) discussed this problem, and adopted a value
for the systemic velocity of $-$7$\pm$10 km s$^{-1}$ (heliocentric
velocities are quoted here, unless noted explicitly).  This value was
inferred from an average of O-type stars in open clusters of the
Carina nebula, of which $\eta$ Car is a member.  Davidson et al.'s
adopted value deserves confirmation, though, because it is not a
direct measurement of $\eta$ Car itself.  For most applications,
typical Doppler velocities are high enough that the uncertainty of
$\pm$10 km s$^{-1}$ is not a severe impediment.  However, the compact
ejecta near the star known as the Weigelt objects (Weigelt \&
Ebersberger 1986) are moving away from the star at only $\sim$50 km
s$^{-1}$; for these slow condensations the $\pm$10 km s$^{-1}$
uncertainty in the systemic velocity is relatively large.

Near-infrared (IR) emission lines of molecular hydrogen arising in the
polar lobes of the Homunculus (Smith 2002) provide an opportunity to
measure the systemic velocity directly for the first time.  Using new
high-resolution data for the H$_2$ $\lambda$21218 emission line, the
present study confirms the accuracy of Davidson et al.'s adopted value
to within about 1 km s$^{-1}$, and improves the precision by a factor
of 10.  Observations are presented in \S 2, the measurement of the
systemic velocity is discussed in \S 3, and \S 4 presents new
high-resolution measurements for the Weigelt knots and discusses
corresponding implications of the systemic velocity.

\section{OBSERVATIONS}

High-resolution ($R\simeq$60,000, $\sim$5 km s$^{-1}$) near-IR spectra
of $\eta$ Car were obtained on 2003 December 12 using the Phoenix
spectrograph on the Gemini South telescope.  Phoenix has a
1024$\times$256 InSb detector with a pixel scale of
0$\farcs$085$\times$1.26 km s$^{-1}$ at a wavelength of $\sim$2
$\mu$m. The 0$\farcs$34-wide long-slit aperture was oriented at
P.A.=310$\arcdeg$ along the polar axis of the Homunculus, and
positioned on the bright central star (identical to the slit position
for the central star used by Smith 2002).  Sky conditions were
photometric, and the average seeing was roughly 0$\farcs$5 (although
the resulting spatial resolution was better for short
exposures of the central star).  Sky chopping was accomplished with an
observation of an off-source position roughly 35$\arcsec$ to the
southeast.

A pair of 60-second exposures sampled the $v$=1$-$0 S(1) emission line
of molecular hydrogen at 21218 \AA \ in the Homunculus, while the
brightest pixels on the central star were saturated.  These saturated
pixels were filled-in using shorter 3-second exposures. This is only
important for determining accurate positional offsets, since Smith
(2002) showed that H$_2$ lines are absent in the spectra of the star
and the Weigelt knots. Matching the wings of the point-spread function
in the saturated data with those in the short exposures allowed the
position of the central star to be determined to better than one pixel
(0$\farcs$085). Figure 1(a) shows long-slit data for the H$_2$ line,
where the reflected continuum light in the Homunculus nebula has been
subtracted out.  Emission from H$_2$ in the front side of the NW polar
lobe and the far side of the SE polar lobe is apparent (see Smith
2002).

The bright standard HR~3685 was observed on the same night with the
same grating settings in order to correct for telluric absorption.
Numerous telluric lines were also used for wavelength calibration,
using the telluric spectrum available from NOAO.  Velocities were
calculated adopting a vacuum rest wavelength of 21218.356 \AA \ for
the H$_2$ 1$-$0 S(1) line (Bragg, Brault, \& Smith 1982), and these
velocities were corrected to a heliocentric reference frame assuming
an adjustment of 13.8 km s$^{-1}$ for the motion of the Earth.
(Heliocentric velocities will be quoted here; for $\eta$ Car, LSR
velocities are offset from heliocentric by $-$11.6 km s$^{-1}$.)
Uncertainty in the resulting velocities is roughly $\pm$1 km s$^{-1}$,
dominated by scatter in the dispersion solution for numerous telluric
lines across the observed wavelength range.

Similar Phoenix spectra of the [Fe~{\sc ii}] $\lambda$16435 emission
line in $\eta$ Car were obtained on 2003 December 10, also on Gemini
South, with the long-slit aperture having the same positon angle
centered on the bright star.  Velocities were calculated the same way,
adopting a vacuum rest wavelength of 16439.981 \AA \ for this line.
The [Fe~{\sc ii}] line is useful with regard to some consequences of
the systemic velocity for the Weigelt knots as discussed in \S 4.  The
observed [Fe~{\sc ii}] line profile is shown in Figure 2, where the
integrated line flux is 5.2$\times$10$^{-10}$ erg s$^{-1}$ cm$^{-2}$.

\begin{table}
\caption{H$_2$ Velocity Measurements}
\begin{tabular}{@{}lccc}       \hline\hline
$\vert R \vert$ &$V_{\rm NW}$  &$V_{\rm SE}$  &$V_{\rm cen}$  \\
(arcsec)        &(km s$^{-1}$) &(km s$^{-1}$) &(km s$^{-1}$)  \\
\hline
2.0             &-72.6         &56.0          &-8.3           \\
2.3             &-70.3         &53.9          &-8.2           \\
3.1             &-53.9         &38.3          &-7.8           \\
\hline
\end{tabular}
\end{table}

\begin{figure}\begin{center}
\epsfig{file=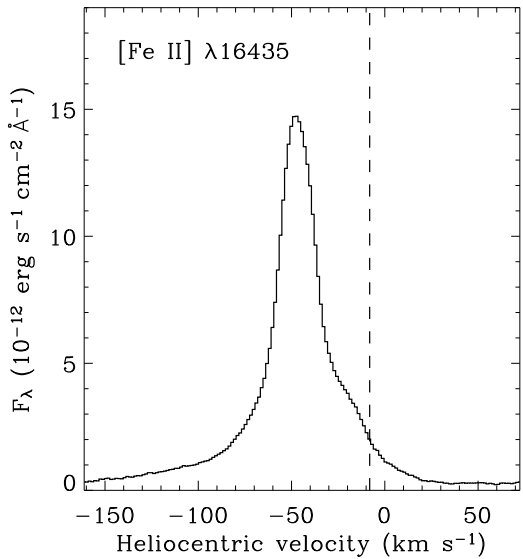,width=3.2in}\end{center}
\caption{Continuum-subtracted [Fe~{\sc ii}] $\lambda$16435 emission
from the Weigelt objects.  This line profile was measured in a
0$\farcs$43-wide segment of the slit centered 0$\farcs$3 northwest of
the star.  The measured value for $V_{\rm sys}$=$-$8.1 km s$^{-1}$ is
marked with a vertical dashed line.}
\end{figure}

\section{THE SYSTEMIC VELOCITY}

Near-IR spectra with lower resolution showed that the H$_2$ line is
emitted {\it exclusively} in the polar lobes (Smith 2002), so that
confusing velocity components from the little homunculus (Ishibashi et
al.\ 2003) and velocity components from the star and Weigelt knots
seen in reflected light in the dusty polar lobes (Smith et al.\ 2003a)
can be avoided.  Smith et al.\ (2003b) also hypothesized that
molecular hydrogen is coincident with a thin layer of cool dust in the
outer parts of the polar lobes, which contains most of the mass in the
Homunculus.  This makes the 21218 \AA \ line an excellent tracer of
the center-of-mass velocity of the Homunculus.  It is a better tracer
than most visual-wavelength lines; while H$\alpha$ shows a narrow peak
at low heliocentric velocity (Boumis et al.\ 1998), it is not clear
where this emission originates (Smith et al.\ 2000) or if it is moving
with respect to the star.  H$_2$ avoids this ambiguity.  Since the
Homunculus was ejected during the Great Eruption, an event which
lasted about 20 years (longer than the 5.5 yr cycle of the putative
binary system), it seems straightforward to assume that H$_2$ also
represents the center-of-mass of the $\eta$ Car system.\footnote{The
only caveat here would be if the Homunculus departs from perfect axial
symmetry.  Morse et al.\ (1998) noted a slight ``banana'' shape to the
Homunculus, which is apparent when an image of $\eta$ Car is viewed
upside-down.  Such a departure from axial symmetry would be manifested
in the data presented here as a systematic change in the centroid
velocity with increasing spatial offset from the star.  This is not
seen within the uncertainty of the velocity measurements, so the
effect is neglected here.}

H$_2$ emission from the front side of the NW polar lobe and the far
side of the SE polar lobe are apparent in Figure 1.  Tracings of the
spectrum were made in 3-pixel (0$\farcs$26) wide spatial segments at
three different offset positions on either side of the star at
2$\farcs$0, 2$\farcs$3, and 3$\farcs$1 from the star (listed as $\vert
R \vert$ in Table 1).  Example tracings at 2$\farcs$3 in each
direction are shown in Figure 1(b).  Velocities measured from these
spectral tracings are listed in Table 1 as $V_{\rm NW}$ and $V_{\rm
SE}$.  These velocity measurements in Table 1 are an average of a
flux-weighted centroid and a Gaussian fit to the line profile;
differences between these two measurement methods were typically
$\pm$0.2 km s$^{-1}$, and in only two cases approached $\pm$1 km
s$^{-1}$ when the line profiles were somewhat asymmetric, so it is a
relatively unimportant source of uncertainty compared to the absolute
wavelength calibration.  The average of $V_{\rm NW}$ and $V_{\rm SE}$
at each position is taken as the heliocentric velocity of the center
of mass of the Homunculus, $V_{\rm cen}$ in Table 1.  The mean of the
three values for $V_{\rm cen}$ listed in Table 1 gives $-$8.1 km
s$^{-1}$ for the systemic velocity of $\eta$ Carinae, with an
uncertainty of $\pm$1 km s$^{-1}$ due to the wavelength calibration.
Correcting to the local standard of rest, the systemic velocity would
be $V_{\rm LSR}$=$-$19.7$\pm$1 km s$^{-1}$.

\section{CONSEQUENCES: THE WEIGELT KNOTS}

This new measurement of $\eta$ Car's systemic velocity
will be useful in future studies of possible orbital reflex
motion, kinematics and geometry of ejecta, or high-resolution UV
echelle spectra of absorption systems, for example. The remaining
discussion here is limited to immediate consequences regarding the
slow-moving compact Weigelt objects.  Their kinematics, including both
radial velocity and proper motion measurements, have been discussed
previously by several authors (Weigelt et al.\ 1995; Davidson et al.\
1997; Dorland et al.\ 2004; Smith et al.\ 2004), and the results have
been somewhat controversial.  From recent proper-motion measurements
using the {\it Hubble Space Telescope} ({\it HST}), Dorland et al.\
(2004) find an ejection date around 1934, while Smith et al.\ (2004)
find an earlier ejection date around 1908 from similar {\it HST} data.
For Doppler velocities, Smith et al.\ (2004) showed that different
emission lines give different results, with velocities around $-$47 km
s$^{-1}$ for high-excitation UV lines, and slower speeds of roughly
$-$40 km s$^{-1}$ for low-excitation optical lines.

Figure 2 shows the line profile of [Fe~{\sc ii}] $\lambda$16435 in the
Weigelt knots.  This is a tracing from long-slit data where a fit to
the continuum has been subtracted.  The profile shows a bright narrow
line core with a fainter red plateau, perhaps indicating multiple
velocity components.  Even the strong narrow component is resolved,
with a FWHM of $\sim$22 km s$^{-1}$ compared to the spectral
resolution of 5 km s$^{-1}$.  The corresponding velocity width of
about 21 km s$^{-1}$ is too wide to be thermal broadening alone.

A flux-weighted centroid gives $-$46.2 km s$^{-1}$ for the Doppler
velocity of the integrated [Fe~{\sc ii}] emission line.  If instead a
two-component fit is used, Doppler velocities of $-$46.9 and $-$18.1
km s$^{-1}$ are found for the bright and faint components,
respectively.  These multiple velocity components or complex line
shape may shed some light on different velocities for different
emission lines measured by Smith et al.\ (2004).  In particular,
variable strength of the red plateau would make the average centroid
velocity appear to change when observed at lower spectral resolution.

A simple interpretation like attributing the fainter component to knot
B, for example, is not feasible; the red plateau has a spatial
separation from the central star of 0$\farcs$27, which is identical to
that for the brighter narrow component.  Interestingly, this
separation is roughly consistent with the measured positions of C
and/or D in {\it HST} images (Smith et al.\ 2004; Dorland et al.\
2004), but not B.

Instead, a more involved explanation is likely.  Suppose the emission
at various Doppler velocities in Figure 2 is caused by real kinematic
motion; Smith et al.\ (2004) proposed that slower velocities may emit
from the dense neutral cores of the condensations, while faster
blueshifted velocities may arise in material ablated from their outer
layers.  One possible interpretation for the line profile in Figure 2,
then, is that the [Fe~{\sc ii}] line is suppressed by collisional
de-excitation in the slower denser cores of the Weigelt knots, while
the stronger and more blueshifted emission arises in their
lower-density ablated envelopes.  Note that the gas which dominates
the emission of [Fe~{\sc ii}] $\lambda$16435 is already at or above
the critical density for this transition (Smith 2002).  If this
interpretation is correct, then the true representative velocity for
the Weigelt knots would be close to $-$40 km s$^{-1}$ (the center at
1/5 of the peak emission) as suggested by Smith et al.\ (2004), rather
than the value of $-$47 km s$^{-1}$ indicated by the stronger narrow
component.

Correcting these velocities for the systemic velocity, and combining
them with proper motions gives clues to their three-dimensional
trajectories (adopting $i$=42$\arcdeg$ as the inclination angle of the
Homunculus; Smith 2002).  Either of these heliocentric Doppler
velocities prohibit the Weigelt knots from residing exactly in the
equatorial plane of the Homunculus if they are combined with the
proper motion measurements by Smith et al.\ (2004), given the smaller
uncertainties in the new measurement of the systemic velocity
presented here.  Interestingly, the faster speed ($-$47 km s$^{-1}$)
would place the condensations within a few degrees of the equatorial
plane if the proper motions by Dorland et al.\ (2004) are correct.
However, there is no clear reason why the Weigelt knots {\it must} be
in the equatorial plane --- as opposed to residing in the approaching
polar lobe of the little homunculus, for example (the velocities
measured here combined with the proper motions by Smith et al.\ 2004
could be consistent with this interpretation).  Clearly, further study
of the Weigelt knots is worthwile.  In addition to continued proper
motion measurements, it would be useful to examine differences in
radial velocity as a function of ionization, excitation, and
especially critical density.

\smallskip\smallskip\smallskip\smallskip
\noindent {\bf ACKNOWLEDGMENTS}
\smallskip
\scriptsize

\noindent Support was provided by NASA through grant HF-01166.01A from
the Space Telescope Science Institute, which is operated by the
Association of Universities for Research in Astronomy (AURA), Inc.,
under NASA contract NAS 5-26555.  Based on observations obtained at
the Gemini Observatory, which is operated by AURA, under a cooperative
agreement with the NSF on behalf of the Gemini partnership: the
National Science Foundation (US), the Particle Physics and Astronomy
Research Council (UK), the National Research Council (Canada), CONICYT
(Chile), the Australian Research Council (Australia), CNPq (Brazil),
and CONICET (Argentina).  These data were obtained in service
observing mode, and I thank the Gemini staff, particularly Bob Blum,
Ken Hinkle, and Bernadette Rodgers for their assistance.

\label{lastpage}
\end{document}